\documentclass[aps,twocolumn,showpacs,preprintnumbers,amssymb]{revtex4}

\newcommand{\be}{\begin{eqnarray}}
\newcommand{\ee}{\end{eqnarray}}

\newcommand{\eins}{\mbox{$1 \hspace{-1.0mm}  {\bf l}$}}
\newcommand{\ket}[1]{\left|{#1}\right\rangle}

\newcommand{\ketbra}[2]{\left|{#1}\rangle\!\langle{#2}\right|}
\newcommand{\mean}[1]{\langle{#1}\rangle}\usepackage{dcolumn} 
\def\eea{\end{eqnarray}}
\def\C{\hbox{$\mit I$\kern-.7em$\mit C$}}
\def\N{\hbox{$\mit I$\kern-.3em$\mit N$}}

\def\tr{{\rm tr}}

\usepackage{bm} 

\usepackage{amsmath,amsfonts,amssymb,graphics,graphicx,epsfig,color,times,bbm}

\begin{document}



\title{Spin gases as microscopic models for non-Markovian decoherence}

\author{L. Hartmann$^{1}$, J. Calsamiglia$^{1}$, W. D{\"u}r$^{2}$ and H.-J. Briegel$^{1,2}$}

\affiliation{$^1$ Institut f{\"u}r Theoretische Physik, Universit{\"a}t Innsbruck,
Technikerstra{\ss}e 25, A-6020 Innsbruck, Austria\\
$^2$ Institut f\"ur Quantenoptik und Quanteninformation der \"Osterreichischen Akademie der Wissenschaften, Innsbruck, Austria.}
\date{\today}

\begin{abstract}
We analyze a microscopic decoherence model in which the total system is
described as a spin gas. A spin gas consists of $N$ classically moving
particles with additional, interacting quantum degrees of freedom (e.g. spins).
For various multipartite entangled probe states, we analyze the decoherence
induced by interactions between the probe- and environmental spins in 
such spin gases.  We can treat mesoscopic environments ($\approx 10^5$ particles). 
  We present results for a lattice gas, which could be realized by
neutral atoms hopping in an optical lattice, and show the effects of non-Markovian and correlated noise, as well as finite size effects. 

\end{abstract}


\maketitle


Decoherence is a fundamental mechanism believed to be responsible for the transition
from the quantum to the classical world \cite{Zu03}. Interactions between system degrees of freedom
and (uncontrollable) environment degrees of freedom lead to entanglement,
manifesting itself in the decoherence of the system state. Many decoherence
models have been discussed in the literature, most prominent among
them oscillator bath models \cite{oscillatorbath} and spin bath models
\cite{Pr00}. Here, we study a physically motivated model of a mesoscopic
inhomogeneous spin bath. In particular, we describe the joint state of system and environment
by a {\em spin gas} \cite{Ca05}. A  spin gas is a system of quantum spins with stochastic time-dependent interactions.  A physical model of a spin gas is
a system of $N$ classically moving particles with additional, internal spin
degrees of freedom. Upon collision, these
quantum degrees of freedom interact according to some
specified Hamiltonian. Hence, in such spin gases, classical kinematics
drives the evolution of the quantum state, and also the decoherence of probe
systems. In general, multiple non-consecutive collisions of particles are possible. 
In this sense, a spin gas provides a microscopic model for {\em non-Markovian decoherence}. 

In this letter, we determine an effective, time-dependent map that describes the
decoherence process in a spin gas. 
For specific system--environment interactions, we can treat
mesoscopic environments ($\approx 10^5$ particles) exactly and efficiently.
 To derive the map, we extend
the description of certain states in terms of Valence Bond Solids (VBS)
\cite{Ve04} to completely positive maps. We do not restrict
the analysis to single-qubit probes, but also consider the effect of
decoherence on various (multipartite) entangled probe states (cf.
\cite{Ca04}). Due to the stochastic nature of the interactions, our model does
not display symmetries, which otherwise simplify the treatment (see e.g.
\cite{Zi04} for homogeneous system--environment interactions). Throughout the
paper, we concentrate on a specific realization of a spin gas, the spin lattice
gas. However, our methods can be easily applied to decoherence in other spin
gases, such as a Boltzmann gas \cite{Ca05}. Spin gases are not only
toy models of theoretical interest, but could be experimentally realized with
present-day technology and existing setups \cite{expref}.

{\bf The model:}
The probe system $A$ consists of $N_A$ qubits prepared in some arbitrary
state. The qubits of system $A$ interact with uncontrollable degrees
of freedom of an environment $B$, leading to decoherence. We consider a {\em
microscopic decoherence model} where the environment is described by a spin
gas. The internal quantum degrees of
freedom of this gas interact according to the time-dependent Hamiltonian
\be
\label{H}
H(t)=\sum_{k<l} g({\bm r}_k(t),{\bm r}_l(t)) H_{kl}. 
\ee
The function $g$ depends on the physical nature of the pairwise spin
interaction described by the Hamiltonian $H_{kl}$.  For such systems, we have
shown \cite{Du05,Ca05} that one can efficiently compute reduced
density operators of up to ten particles even for mesoscopic system sizes
($\approx 10^5$ particles), if all Hamiltonians $H_{kl}$ commute and the
initial state of the system is a pure product state. Here, we extend these
results to take arbitrary initial system states and mixed 
environmental states into account, and thereby study the decoherence of
multipartite probe states in a mesoscopic environment efficiently and exactly.

For commuting Hamiltonians $H_{kl}$, the joint state of system and environment
at a time $t$, $|\Psi_{t}\rangle = \prod_{k,l} e^{-i\varphi_{kl}(t)H_{kl}}
|\Psi_0\rangle$, is determined by $N(N-1)/2$ interaction phases
$\varphi_{kl}(t) \equiv \int_0^t g({\bm x}_k(t'),{\bm x}_l(t'))dt'$. These
phases are associated with the adjacency matrix $\Gamma(t)$ of a weighted graph
$G$. The matrix element $\Gamma_{kl}(t)=\varphi_{kl}(t)$ describes the neighborhood relation
of particles $k$ and $l$, or, equivalently, the interaction history. We are interested
in the state of system $A$, i.e. the reduced density operator
$\rho_A(t)=\tr_B|\Psi_t\rangle\langle\Psi_t|$. The commutation of the interaction Hamiltonians greatly simplifies the computation of this operator: 
To compute this operator we first take into account only interactions between particles $k \in A$ and $l \in B$, since interactions within $B$ do not change the state of system $A$. In contrast to the general case of non-commuting Hamiltonians \cite{Da04}, 
entanglement in the environment does not influence the decoherence properties of the system. On top of that, the interactions within
system $A$ itself can be applied to the resulting state afterwards.

In the following, we treat the case $H_{kl}=|11\rangle_{kl}\langle 11|$
and initially completely polarized environment spins,
$|\Psi\rangle_B=|+\rangle^{\otimes N}$ where $|\pm\rangle\!\propto\! |0\rangle
\pm|1\rangle$. Extensions to other commuting
interaction Hamiltonians and arbitrary (possibly mixed) product environmental
states are straightforward. The isomorphism between completely positive maps and
mixed states \cite{Ci00} ---together with a generalized Valence Bond Solids
(VBS) picture \cite{Ve04,Du05}--- determines the
effective map ${\cal E}_t$ that takes an initial state at $t_0=0$ to the
state at time $t$, i.e. $\rho_{A}(t) ={\cal
E}_t \rho_A(0)$. This map ${\cal E}_t$ can be equivalently described by the state
$E_t= \eins^{A'}\otimes {\cal E}_t^{A}
|\Phi\rangle\langle\Phi|$. Here, $|\Phi\rangle =
\otimes_{k=1}^{|{A}|}|\phi^+\rangle_{k'k}$, $|\phi^+\rangle
\!\propto\!|00\rangle + |11\rangle$, and 
${A'}$ is an auxiliary system with the same dimension as $A$. We can express the map in the Pauli basis
$\sigma_k$, where $\sigma_0\equiv \eins$. Then, ${\cal E}_t\rho = \sum
\lambda_{k_1\ldots k_{N_A},l_1\ldots l_{N_A}} \sigma_{k_1}\ldots \sigma_{k_{N_A}}
\rho \sigma_{l_1}\ldots \sigma_{l_{N_A}}$, where $\lambda_{k_1\ldots
k_{N_A},l_1\ldots l_{N_A}} = \langle \phi_{k_1\ldots k_{N_A}}|E_t|\phi_{l_1\ldots
l_{N_A}}\rangle$ with $|\phi_{k_1\ldots k_{N_A}}\rangle=\eins^{\bm A'} \otimes
(\sigma_{k_1}\ldots \sigma_{k_{N_A}})^{A} |\Phi\rangle$. The coefficients of
the map are given by the coefficients of the state $E_t$ written in (tensor
products of) Bell bases. As in the case of states \cite{Du05}, we can 
separately consider maps (or equivalently the states $E_t^{(l)}$) resulting
from the interaction of the system with a single particle $l$ in the
environment. 
We find $E_t^{(l)}=1/2(|\Phi\rangle\langle \Phi| + \otimes_{k=1}^{|{A}|}
|\chi_k\rangle\langle\chi_k|)$ with $|\chi_k\rangle_{k'k}=1/\sqrt{2}(|00\rangle
+ e^{-i \varphi_{kl}(t)}|11\rangle)$, where ${\varphi_{kl}}(t)$ is the
effective interaction phase between particles $k \in {A}$ and $l\in {B}$. The
state $E_t$ describing the total decoherence process incorporates the influence of all
particles $l \in {B}$. We obtain $E_t$ (up to normalization) by calculating the
Hadamard product of all $E_t^{(l)}$ written in the standard
basis, i.e. by component-wise multiplication. The matrix elements of $E_t$
expressed in the tensor Bell basis finally determine ${\cal E}_t$. We find that $E_t$ has
non-zero components only in the subspace spanned by $\{|\phi_{k_1\ldots
k_{N_A}}\rangle\}$ with $k_j \in \{0,3\}$. The map ${\cal E}$ thus
contains only tensor products of Pauli operators $\eins$ and $\sigma_z$. 
Equivalently, we can express the action of the map on an arbitrary probe input
state $\rho=\sum_{s,s^\prime} \rho_{{\mathbf s}{\mathbf s'}}(0) |{\mathbf s}\rangle\langle
{\mathbf s'}|$ by determining the evolution of the coherences $\rho_{{\mathbf
s}{\mathbf s'}}(t)$. With ${\mathbf s_A},{\mathbf s'_A}$ we denote binary vectors
of length $N_A$. We can express the coherences as $\rho_{{\mathbf s}{\mathbf s'}}(t)= C_{{\mathbf
s}{\mathbf s'}}(t) \rho_{{\mathbf s}{\mathbf s'}}(0)$ with \cite{Ca05}
\be
C_{{\mathbf s}{\mathbf s'}}(t)
=e^{i
\frac{1}{2}\sum_k(\mathbf{s}_A-\mathbf{s}_A')\cdot\mathbf{\Gamma_{k}}}
\prod_{k=1}^{N_B}
\cos[\textstyle{\frac{1}{2}}(\mathbf{s}_A-\mathbf{s}_A')\cdot\mathbf{\Gamma_{k}}].
\ee
The $(\mathbf{\Gamma_{k}})_j=\Gamma_{k j}$ for each particle $k\in B$ are
$N_A$-dimensional vectors. The method described above can be considered as an
extension of the generalized VBS picture for states to one for completely
positive maps. In this picture, we can determine the evolution of arbitrary system
states in a mesoscopic spin environment, since the computational effort to
calculate the maps scales only linearly with the number of
particles in the environment (as opposed to exponentially for general
system-environment interactions). The size of the
probe system is
limited to about 10 spins for numerical computations due to exponential scaling with system size $N_A$. 

The quantum properties of the system are directly linked to the classical
statistical properties of the gas through $\Gamma(t)$. In general, it is
thus necessary to know the classical $n$-body phase-space distributions to give a complete
description of the quantum state. If we assume no
control over quantum or classical degrees of freedom of the background gas we
should average over all possible collision patterns at any given time \cite{Ca05}:
$\bar{C}_{s,s'}(t)=\int \mathrm{d}\Gamma\, p_t(\Gamma) C_{s,s'}(\Gamma)$,
where $p_t(\Gamma)$ is the probability that at time $t$ the interaction history is given by $\Gamma$. 

For some gas models and regimes (like the Boltzmann gas studied in \cite{Ca05}), correlations play a minor role and one can find analytical expressions for single-particle phase-space
distributions.  In this paper,  however, we study  a lattice gas model
that exhibits strong correlations, and produce the different random
realizations of $\Gamma(t)$ by direct simulation of the gas.

The lattice model can be possibly implemented in a
quantum optical system. It has already been demonstrated that an optical
lattice can be used to store ultra-cold atomic gases. The degree of control in
these experiments is extraordinary, opening the door to a wide range of experiments and
theoretical proposals \cite{greiner02,mandel03,paredes04,jaksch98,hofstetter02,duan03,damski03}. One can choose a parameter regime where each lattice site is
occupied by at most one atom \cite{jaksch98,greiner02}. The internal
state of  the atom (e.g. two meta-stable hyperfine states) can be stored in
coherent superpositions over long time-scales (few minutes). Coherent
inter-atomic interactions have been achieved by cold collisions
\cite{mandel03}. These correspond precisely to the Ising-type
interactions chosen here. One can also find schemes \cite{expref} to induce a
random (incoherent) hopping of atoms from one site to its neighboring
sites. Hence, we consider an $M\!\times\!M$ lattice containing $N$ particles
that  randomly hop from site to site with a hopping rate  $\eta$, and have
nearest-neighbor interactions with coupling constant $g_o$. With special relevance to possible experiments, we note that all results in this paper hold even when the environment particles themselves decohere. The only requirement is that the diagonal elements  of the environment's state ---in the canonical  basis--- remain unchanged.

{\bf Decoherence of a single qubit: }
For a system consisting of a single qubit ${A}=\{1\}$
and an arbitrary environment, the time dependent map corresponding  to a particular collisional history is
\be
{\cal E}_t\rho =\lambda_{00}\rho + \lambda_{11} \sigma_z\rho\sigma_z +
\lambda_{01}(\eins\rho\sigma_z - \sigma_z\rho\eins), \label{eq:sqmap}
\ee
with $\lambda_{00}=(1+r\cos \gamma)/2$, $\lambda_{11}=(1-r\cos\gamma)/2$, and
$\lambda_{01}=(i r\sin\gamma)/2$ where $r(t)=\prod_{l\in {B}}
\cos(\frac{\varphi_{1l}(t)}{2})$, $\gamma(t)=\sum_{l\in {B}}
\frac{\varphi_{1l}}{2}$. 
Depending on the parameter regime, semi-quantal gases can follow various
collision patterns. Accordingly, the dynamics of their quantum properties can
differ considerably.  If in every time step $\delta t$ a given particle collides with a different
particle and acquires an interaction phase $\delta_\varphi $, the dynamics will
be purely Markovian. The coherence of that particle will decay exponentially
fast with the number of time steps $k=\Delta t /\delta t$, 
$|\rho_{01}|=[\cos(\delta_\varphi/2)]^k=e^{-\Delta t/\tau_e }$ with
$\tau_e\approx 8 \delta t/\delta_\varphi^{2}$.  If, on the other hand,  in a small
time interval $\Delta t$ a given gas particle has collided $k$ times with
the same particle, the coherent addition of the interaction phase leads to a
Gaussian type of decay:   $|\rho_{01}|=\cos(k \delta_\varphi/2)\approx e^{-
\Delta t^2/(2\tau_g^2)}$ with $\tau_g=2 \delta t/\delta_\varphi$.
The exponential and the Gaussian decay are the two extreme cases, the
dynamics of the coherence $|\rho_{01}|$ will usually lie in between. This
also holds for the coherences in a multi-qubit density matrix. 
A complete characterization of the system's decoherence is obtained by averaging the maps ${\cal E}_t$ over the possible collision patterns. The resulting map has the same form as \eqref{eq:sqmap}, but replacing the coefficients $\lambda_{ij}$ by their average values. 
An analytical expression for the time dependence of these averaged coefficients, and hence of the decoherence process, is in general hard to obtain. However, for finite lattices, one can give a precise  description of the time dependence.

{\bf Decoherence of bipartite entangled states: }
We now study the decoherence of different, initially entangled
two-qubit states. In order to explore different regimes, we imagine a scenario
where the probe particles can be displaced at a constant speed $v$ relative to the gas.
 By varying the speed $v$ and the distance $d$ between the probe particles, we can highlight two effects: 
(i) By decreasing the probe speed, we analyze the effect of multiple
interactions with the same particle in contrast to interactions with different
(independent) particles.
(ii)  By increasing the distance, we turn from correlated to independent
collisions between probe and environment particles.%
\begin{figure}[ht]
\begin{picture}(210,58)
\put(-20,-10){\includegraphics[width=120pt]{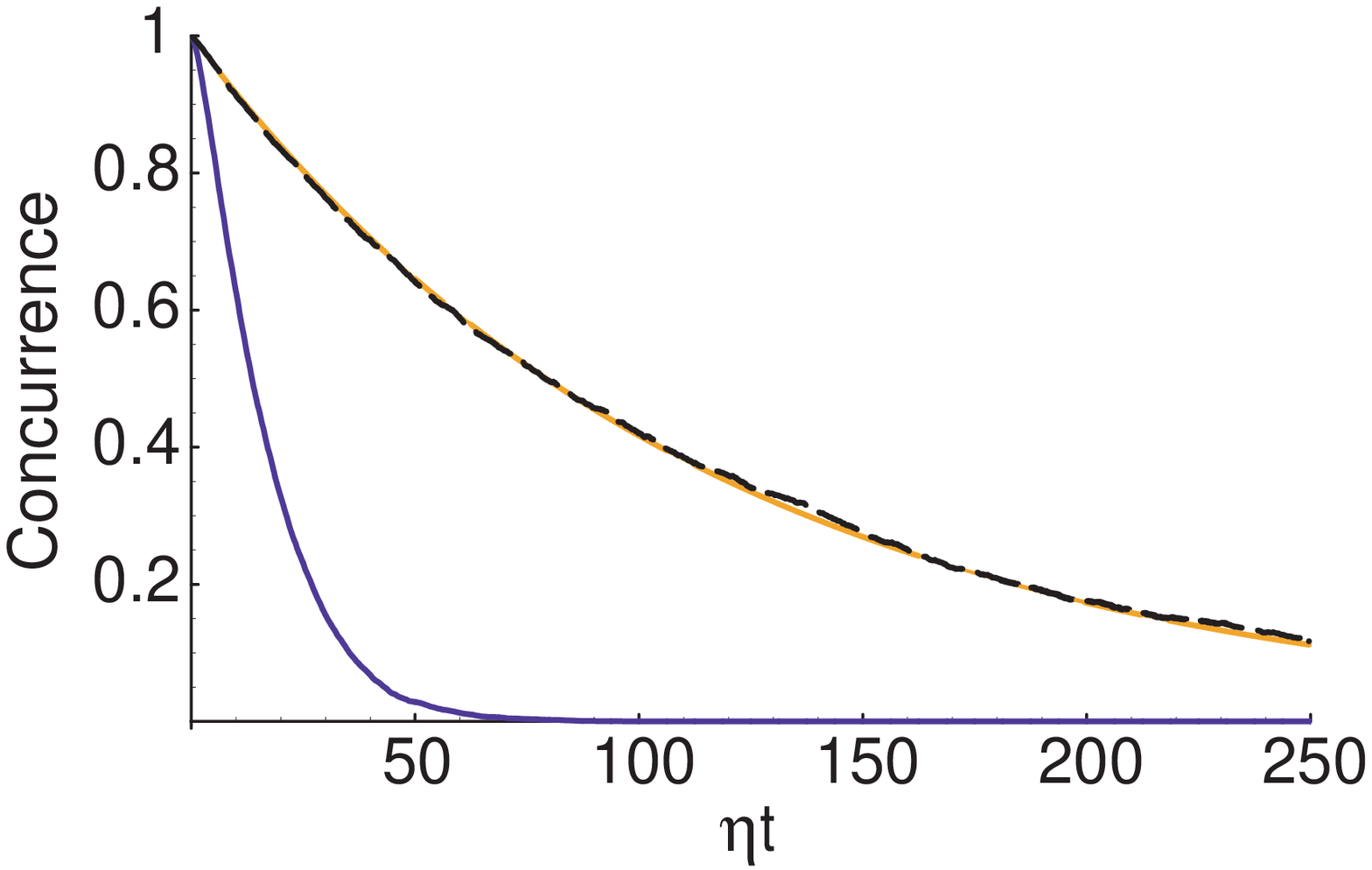}}
\put(85,-12){\includegraphics[width=140pt]{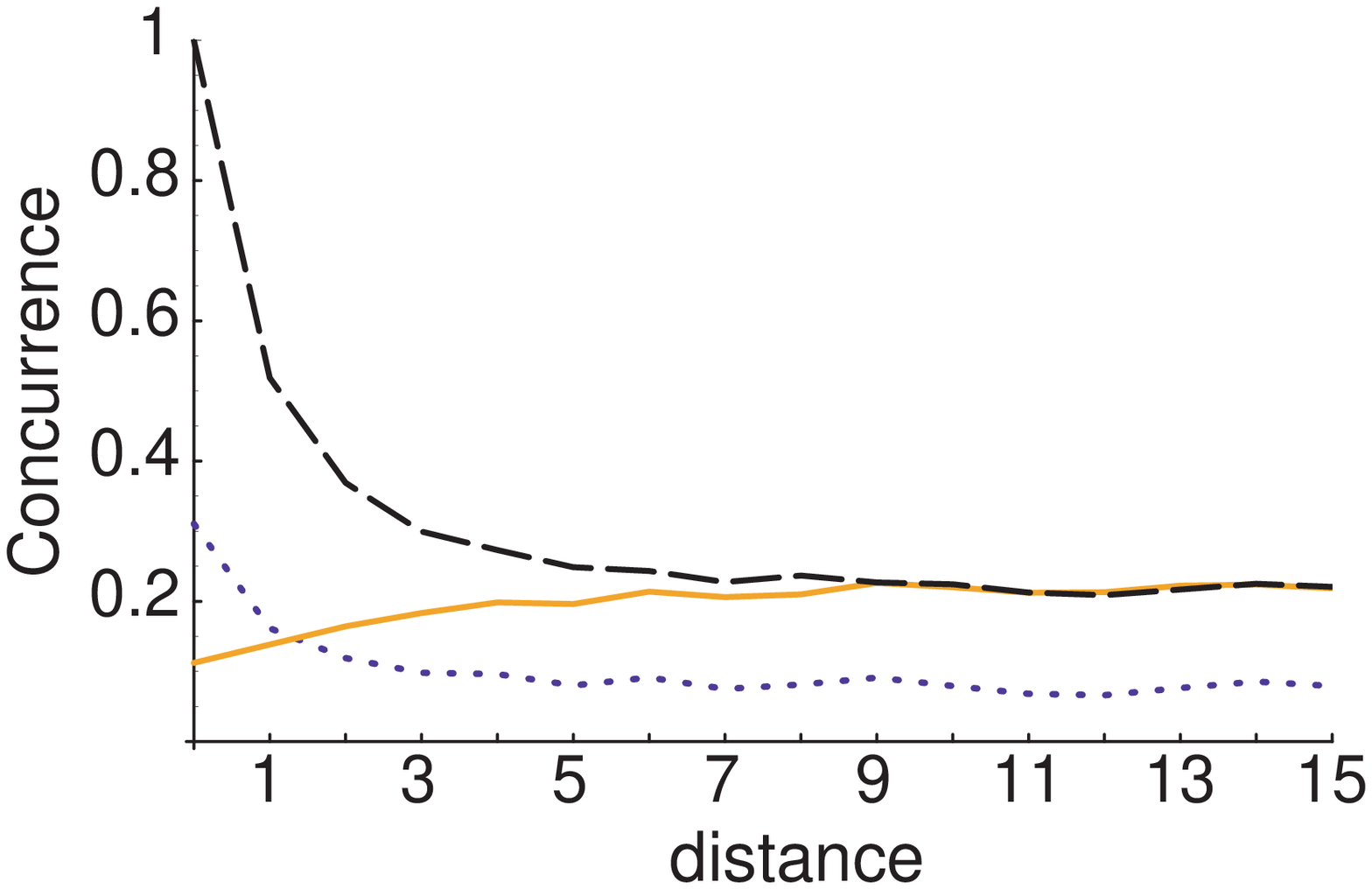}}
\put(50,54){(a)}
\put(180,54){(b)}
\end{picture}
\caption[]{\label{fig:decoh} (a) Concurrence of two probe particles prepared in
a maximally entangled Bell state $\ket{\phi^+}\!\propto\! \ket{00}+\ket{11}$ 
in a  $100\!\times\!3200$ lattice gas  with $N=8\times 10^4$  for  $g_o=0.8
\eta $ and fixed probes (solid);  and fast moving probes (dashed). The solid,
light curve corresponds to an analytical result obtained in a Markovian regime.  (b)
Concurrence of two probe particles at  time $t_1=25 \eta^{-1}$ for different
initial entangled states as a function of distance $d$ between the probes :
$\ket{\psi^+}\!\propto\! \ket{01}+\ket{10}$ (dashed-dark) , $\ket{\phi^+}$
(solid-light), and $\ket{G}\!\propto\! \ket{+0}+\ket{-1}$ (dotted). }
\end{figure}

Figure \ref{fig:decoh}(a) shows the decay of entanglement, measured by the
concurrence \cite{wootters98}, of an initial Bell state
$\ket{\phi^+}$ in two extreme
scenarios: (i) Fixed probe particles ($v=0$). (ii) Large probe speeds, $v/a\gg \eta$, where $a$
is the inter-site spacing. A fixed value $\varphi=0.1$ is assigned 
to the collisional phase every time a probe particle crosses an occupied site.
These two scenarios  illustrate the difference between Markovian and
non-Markovian environments. A large probe speed enforces a perfect Markovian
behavior which matches the analytical curve \footnote{At every time step,
each probe particle interacts with a new environment particle
with probability $\nu$. Hence, after a number $s$ of time steps, the relevant
coherence  is given by $|C_{00,11}|=|C_{0,1}|^2=| \nu \exp(i \delta_\varphi/2)
\cos(\delta_\varphi/2)+(1-\nu)|^{2 s}$.}. Figure \ref{fig:decoh}(b) shows the
concurrence at a given time $t_1$ as a function of the distance for three
different entangled states: two Bell states and a cluster state (see figure
caption). For Bell states  the concurrence is equal to the absolute value of
their only non-zero off-diagonal element in the density matrix, and therefore
Figure \ref{fig:decoh} provides direct information about the individual
coherences. The figure shows the influence of correlated collisions:
coherences $\rho_{01,10}$ (in $\ket{\psi^+}$) are robust against
correlated noise, and coherences $\rho_{00,11}$ (in $\ket{\phi^+}$) are
especially fragile under correlated noise. The remaining coherences decay
in the same way under correlated or uncorrelated noise (hence, the weak
distance dependence of $\ket{G}$). From Fig. \ref{fig:decoh}(b)  we also see that
the immediate environments of each probe become more independent
as $d$ increases.

The different behavior under correlated and uncorrelated collisions can be
readily understood. For two probe particles ($1$ and $2$) with  very  similar
collision patterns, i.e., $\Gamma_{2j}= \!\Gamma_{1j}\!+\!\delta_j$ for all $j$,
coherences associated with $\ketbra{01}{10}$ will only decay by a factor
$2^{-N_B} \sum_{s_B}e^{i\mathbf{\delta}\cdot\mathbf{s}_B}$, while
$\ketbra{11}{00}$  will be ``super-damped'' by $2^{-N_B} \sum_{s_B}e^{i (2
\mathbf{\Gamma_1}+\mathbf{\delta})\cdot\mathbf{s}_B}$. Therefore,
classical correlations in the collisions, e.g. induced by the geometry of the
set-up, can significantly influence the entanglement properties of the system.

{\bf Decoherence of multipartite entangled states: }
We now apply our method to investigate the decoherence of different
multipartite entangled probe states of $N_A$ qubits. Due to the lack of simple,
computable multipartite entanglement measures, we use the negativity 
of bipartitions \cite{Vi02} as an indicator of multipartite entanglement in the system.
That means, we consider bipartitions of the system, i.e. a partition consisting of a set of particles $A_{k}$ and its complement $\bar{A}_{k}$,   and
investigate the entanglement properties with respect to the $2^{N_A-1}-1$
independent bipartitions \cite{Du00}. In general, we get a broad
picture of multipartite entanglement in this way. 
For each bipartition  we can determine its negativity ${\cal N}_{A_{k}}=(||\rho^{T_{A_{k}}}||_1-1)/2$  \cite{Vi02}.
We define two multipartite entanglement measures: (i) the {\em average negativity}
${\bar {\cal N}}$, as the average over all bipartitions, $\bar{\cal N} = 1/(2^{(N_A-1)}-1)\sum_{A_{k}} {\cal N}_{A_{k}}$, and (ii) ${\cal N}_{min}=\min\{ {\cal N}_{A_k}\}$.
Zero average negativity is a necessary condition for full separability of the
state, and ${\cal N}_{min}=0$ is a sufficient condition that
the state is not multi-party distillably entangled.

\begin{figure}[ht]
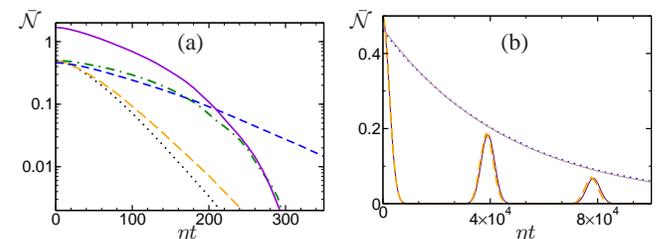

\setlength{\unitlength}{1cm}
\begin{picture}(8,2.5)
\put(-0.17,-0.13){\includegraphics[scale=0.170,trim= 1 0 0 1,clip=true]{Negs.eps}}
\put(4.3,-.13){\includegraphics[scale=0.170,trim= 1 0 0 1,clip=true]{FSRev.eps}}
\put(-0.2,2.5){\footnotesize $\bar {\cal N}$}
\put(1.9,-0.3){\footnotesize $\eta t$}
\put(6.4,-0.3){\footnotesize $\eta t$}
\put(4.2,2.5){\footnotesize $\bar{{\cal N}}$}
\put(1.9,2.2){(a)}
\put(6.2,2.2){(b)}
\end{picture}
\caption[]{\label{Negdecay} (a) Log-linear plot showing the decay of the average negativity $\bar {\cal N}$ for six-qubit linear cluster state (solid), W-state (short-dashed), and GHZ states
$|\Psi\rangle$ (dotted), $|\Psi'\rangle$ (dash-dotted), and $|\Psi''\rangle$
(long-dashed) in a lattice gas with 400 spins in a $40\times 40$ lattice and coupling constant
$g_0=0.8\eta$. (b) Entanglement decay of four-qubit GHZ (long-dashed) and W (dotted) states in  $20\times 20$ lattice with  25 spins and $g_0=8\times 10^{-3}\eta$ illustrating finite size effects. Single-parameter ($\alpha's$) analytical fits are also shown (thin lines).}
\end{figure}

We have examined different multipartite entangled probe states that interact
with a lattice gas through a pairwise Hamiltonian $H_{kl}=\sigma_z^{(k)}\otimes
\sigma_z^{(l)}$. These states are (linear)
cluster states $|\chi\rangle$ \cite{Ra01}, GHZ-states and W-states $|W\rangle\!\propto\!\sum_{i=1}^{N_A} \ket{w_i}$, where $\ket{w_i}$ is the state corresponding to all spins in state zero except the $i^{th}$ that is in one. Since the noise process is basis dependent, we study variants of GHZ
states corresponding to different local bases: 
$|\Psi\rangle \!\propto\! |0\rangle^{\otimes N_A} + |1\rangle^{\otimes N_A}$, 
$|\Psi'\rangle\!\propto\! |0\rangle|+\rangle^{\otimes N_A-1} + |1\rangle|-\rangle^{\otimes N_A-1}$, 
$|\Psi''\rangle\!\propto\! |+\rangle|0\rangle^{\otimes N_A-1} +
|-\rangle|1\rangle^{\otimes N_A-1}$. The decay of the average
negativity is plotted in Fig. \ref{Negdecay}(a). To qualitatively understand
the different behavior of the curves we first derive analytic results
in the limit of independent environments for each particle $k \in {A}$. In this
limit, the decoherence process can be described by a tensor product of single-qubit dephasing maps
${\cal E}_t^{(k)}\rho = p_k\rho + (1-p_k)\sigma_z^{(k)}\rho \sigma_z^{(k)}$,
where $p_k = \frac{1}{2}(1+\prod_{l \in B} \cos(2\varphi_{kl}(t)))$.
Coherences decay as
$\rho_{01}^{(k)}(t)= (2p_k-1) \rho_{01}(0)$.
 As before, the precise time dependence of decoherence will be given by the average $p(t)=\mean{p_k}_{\Gamma(t)}$ over different realizations. We assume that this average value is the same for every probe particle $k$.
Under the action of this dephasing map the family of GHZ states remain diagonal in the GHZ-type basis \cite{Du00}. For such states, we can obtain the spectrum of the partial transposed operators with respect to any bipartition analytically~\cite{Du04}, and calculate average negativities for the three GHZ states given above.  
Here we give those with simple expressions:
$\bar{\cal N}= \frac{1}{2}|2p-1|^{N_A}$, 
and $\bar{\cal
N}^{\prime\prime}=\frac{1}{2^{N_A-1}-1}[{\cal
N}_{min}^{\prime\prime}+(2^{N_A-1}-2)\frac{1}{2}|2p-1|^{N_A-1}]$ with 
${\cal N}_{min}^{\prime\prime}=\frac{1}{2}\max\{0,|2p-1|+|2p-1|^{N_A-1}+|2p-1|^{N_A}-1\}$.
For $W$-states, direct calculation leads to  
$\bar{{\cal
N}}^W=\frac{|2p-1|^2}{N_A(2^N_A-2)}\sum_{a=1}^{N_A-1}\binom{N_A}{a}\sqrt{a(N_A-a)}$.

Several observations follow from these analytic results for independent environments:
(i) Standard GHZ,  $|\Psi\rangle$, and $W$ states remain  $N_A$-party distillable for all times, since ${\cal N}_{min}$ only reaches zero asymptotically  as $t\rightarrow \infty$. This does not hold for the  $|\Psi''\rangle$ GHZ, which has one partition that becomes disentangled at finite $t$, nor for $|\Psi'\rangle$, for which $\bar{\cal N}$  vanishes at a finite time (all partitions are disentangled).  In Fig. 2 we see that $\bar{\cal N}$ also vanishes at finite time for the cluster state.
(ii) In the limit of large system sizes, the average negativity is, to a very good approximation, given by the negativity of the half--half partition (the distribution of partitions with $k$ particles is sharply peaked at $k=N_A/2$). (iii) GHZ-type states decay exponentially as we increase the system size $N_A$, while for $W$ the coherences do not vary with the system size, rendering a weak dependence of its average negativity (constant to first order).
(iv)  For the  states studied here, if a partition is initially more entangled than another, it will remain so also at later times. This does not hold in the presence of correlated collisions, which occur  when the distance between probe particles is not large with respect to the relevant times $t$, i.e. $d<\sqrt{\eta t}$.

In finite lattices and after long enough times the above description fails due to inevitable correlated collisions. However, the finite size leads to interesting effects that can also be easily understood. Figure  2(b)  shows one such effect:  the coherence (and hence the probe's entanglement) lost due to the stochastic interaction with environment spins, is partially recovered after a characteristic revival time.
After long times a given environment particle will have collided $n_k$ times with a probe particle $k$. After $s\gg M^2$ steps, we describe the distribution of values $n_k$ by a Gaussian of mean value $\mean{n_k}=n=4 s/M^2$ and the variance $\sigma^2=\mean{(n_k-n)^2}=\alpha n$, where $\alpha$ depends on the particular lattice model.  We expect that at long times most blocking effects between environment particles will be washed out, and therefore assume that different environment particles will have independent collision distributions. Hence, the total effect of $N_B$ on a particular coherence will scale as $C^{N_B}$, where $C$ is the decay factor of the coherence due to a single environment  particle. The value of $C$ is given by an average taken over a Gaussian of mean $\varphi_o$ and width $\sigma_\varphi$: $C=\mean{\cos(2\varphi)}=\cos (2\varphi_o)\exp(-8\sigma_\varphi^2)$.
The phase $\varphi$ is a sum of the collisional phases (with the corresponding signs)  involved in the particular coherence. For example, for a standard GHZ all phases are added $\varphi=\sum_{k=1}^{N_A} \varphi_{1k}$ leading to a mean value $\varphi_o\approx N_A n\delta_\varphi$ and a variance $\sigma_\varphi^2\approx \alpha' N_A  n \delta_\varphi^2$ \footnote{The factor $\alpha'$ (and $\alpha''$ below) also includes the contribution of $K_{kk'}=\mean{n_{k'} n_k}-n^2\!\propto\! n$, which depends on the distance between the probe particles $k,k'$.}. For the $W$-state and a given coherence, say $\rho_{w_1 w_2}$, the phase is $\varphi=\varphi_{1k}-\varphi_{2k}$,  leading to a vanishing mean value and to a purely exponential decay with $\sigma_\varphi^2 \approx\alpha'' 2 n \delta_\varphi^2$  ---which is independent of the system size $N_A$. 
The exponential decay of the $W$-state and the periodic revival of the GHZ state can be clearly identified in Fig. 2(b).

{\bf Summary:} 
We have studied a microscopic, exact model for non-Markovian decoherence where
the joint system is described by a spin gas. Using a generalized VBS picture for maps, we
have determined the time-dependent maps for the decoherence process. We
studied the decay of entanglement for different multipartite entangled probe
states in a lattice gas with a mesoscopic number of particles. Depending on
the parameters of the gas, we have shown how to reach qualitatively different regimes such as Markovian- and non-Markovian, and correlated and non-correlated
decoherence processes. For finite lattices we find that, although the interactions with the environment are stochastic,  entanglement in the probes can spontaneously revive at a time given by the size of the lattice and independent of the number of environment particles.


We thank J. Asb\'oth for reading the manuscript. This work was supported by the FWF, the European Union (IST-2001-38877,-39227, OLAQUI, SCALA), the DFG, and the \"OAW through project APART (W.D.).


\end{document}